\documentclass[preprintnumbers,article,amsmath,amssymb,floatfix,10pt,prd,twocolumn,
superscriptaddress,nofootinbib]{revtex4-2}
\usepackage{bm}
\usepackage{amsfonts}
\usepackage{latexsym}
\usepackage[latin1]{inputenc}
\usepackage{graphicx}
\usepackage{amsmath}
\usepackage{palatino}
\usepackage{mathpazo}
\usepackage{textcomp}
\usepackage{float}
\usepackage{booktabs}
\usepackage{dcolumn}
\usepackage{ragged2e}
\usepackage{hyperref}
\hypersetup{colorlinks,citecolor=blue}
\hypersetup{colorlinks=true,linkcolor=red,filecolor=magenta,    urlcolor=blue}
\usepackage{amsmath}
\usepackage{xcolor}
\usepackage{orcidlink}
\usepackage{epsfig}
\usepackage{caption}
\usepackage{subcaption}
\usepackage{commath}
\def\jnl@style{\it}
\def\aaref@jnl#1{{\jnl@style#1}}

\def\aaref@jnl#1{{\jnl@style#1}}

\def\aj{\aaref@jnl{AJ}}                   
\def\apj{\aaref@jnl{ApJ}}                 
\def\apjl{\aaref@jnl{ApJ}}                
\def\apjs{\aaref@jnl{ApJS}}               
\def\apss{\aaref@jnl{Ap\&SS}}             
\def\aap{\aaref@jnl{A\&A}}                
\def\aapr{\aaref@jnl{A\&A~Rev.}}          
\def\aaps{\aaref@jnl{A\&AS}}              
\def\mnras{\aaref@jnl{Mon.~Not.~Roy.~Astron.~Soc.}}             
\def\prd{\aaref@jnl{Phys.~Rev.~D}}        
\def\prc{\aaref@jnl{Phys.~Rev.~C}}  
\def\prl{\aaref@jnl{Phys.~Rev.~Lett.}}    
\def\qjras{\aaref@jnl{QJRAS}}             
\def\skytel{\aaref@jnl{S\&T}}             
\def\ssr{\aaref@jnl{Space~Sci.~Rev.}}     
\def\zap{\aaref@jnl{ZAp}}                 
\def\nat{\aaref@jnl{Nature}}              
\def\aplett{\aaref@jnl{Astrophys.~Lett.}} 
\def\apspr{\aaref@jnl{Astrophys.~Space~Phys.~Res.}} 
\def\physrep{\aaref@jnl{Phys.~Rep.}}      
\def\physscr{\aaref@jnl{Phys.~Scr}}       
\def\commat{\aaref@jnl{Comm.~Math.~Phys.}}              
\def\science{\aaref@jnl{Science}}               
\def\cqg{\aaref@jnl{Classical Quant.~Grav.}}            
\def\jpcs{\aaref@jnl{JPCS}}                                     
\def\ijmpd{\aaref@jnl{Int.~J.~Mod.~Phys.~D}}                    
\def\grg{\aaref@jnl{Gen.~Relat.~Gravit.}}               
\def\rpp{\aaref@jnl{Rep.~Prog.~Phys.}}          
\def\npa{\aaref@jnl{Nucl.~Phys.~A}}        
\def\lrr{\aaref@jnl{Living Rev.~Rel.}}                   
\def\jcap{\aaref@jnl{J.~Cosmology Astropart.~Phys.}}    
\def\rmp{\aaref@jnl{Rev.~Mod.~Phys.}}   
\def\epjc{\aaref@jnl{Eur.~Phys.~J.~C}} 
\def\plb{\aaref@jnl{~Phy.~Lett.~B}} 
\def\mpla{\aaref@jnl{Mod.~Phy.~Lett.~A}} 
\def\arxiv{\aaref@jnl{arxiv.org}}

\allowdisplaybreaks[1]

\begin{document}
\color{black}       
%
\title{Slow-roll inflation in $f(T,\mathcal{T})$ modified gravity}

\author{Aaqid Bhat\orcidlink{0000-0002-8644-0677}}
\email{aaqid555@gmail.com}
\affiliation{Department of Mathematics, Birla Institute of Technology and
Science-Pilani, Hyderabad Campus, Hyderabad-500078, India.}

\author{Sanjay Mandal\orcidlink{0000-0003-2570-2335}}
\email{sanjaymandal960@gmail.com}
\affiliation{Faculty of Mathematics and Computer Science, Transilvania University, Iuliu Maniu Str. 50, 500091 Brasov, Romania}

\author{P.K. Sahoo\orcidlink{0000-0003-2130-8832}}
\email{pksahoo@hyderabad.bits-pilani.ac.in}
\affiliation{Department of Mathematics, Birla Institute of Technology and
Science-Pilani, Hyderabad Campus, Hyderabad-500078, India.}
\date{\today}

\begin{abstract}\justifying

In this article, we explore the concept of cosmological inflation within the framework of the $f (T,\mathcal{T})$ theory of gravity, where $f$ is a general function of the Torsion scalar $T$  and the trace, $\mathcal{T}$, of the energy-momentum tensor. It is assumed that the conditions of slow-roll inflation are applicable in$f (T,\mathcal{T})$ gravity. To determine different observables related to inflation, such as the tensor-to-scalar ratio  $r$, scalar spectral index  $n_s$, spectral index $ \alpha_s $, and tensor spectral index $n_t$, the Hubble slow-roll parameters are utilized for a particular model of $f (T,\mathcal{T})$. Lastly, an assessment has been carried out to determine the feasibility of the models by conducting a numerical analysis of the parameters. The findings indicate that it is feasible to achieve compatibility with the observational measurements of slow-roll parameters by utilizing different values of the free parameters.\\

\textbf{Keywords:} $f(T,\mathcal{T})$ gravity; early-time cosmology; Dark Energy; equation of state; slow-roll inflation. 

\end{abstract}

\maketitle



\section{Introduction}
Recent observations have provided clear evidence that our universe is expanding at an accelerating rate and is primarily made up of dark matter and dark energy \cite{Riess/1998,Perl/1999,Eisenstein/2005,Percival/2007,Komatsu/2011}. To account for this phenomenon, alternative theories of gravity are being explored that do not rely on unknown sources of energy. Modified gravity models, including $f(R),  f(G), f(R,G)$, and $f(R,T)$  \cite{Staro/2007,Capo/2008,Chiba/2007,Harko/2011,Moraes/2017}, are able to explain the expanding acceleration of the universe. The $f(R)$ model was the initial modified gravity theory, which was introduced by Buchdahl in 1970. In this theory, the curvature scalar $R$ in the standard Einstein-Hilbert action is replaced by an arbitrary function of $f(R)$. Modified gravitational theories typically involve expanding upon the Einstein-Hilbert action in General Relativity, which is based on the curvature description of gravity. However, an intriguing alternative arises when we extend the action of the equivalent formulation of GR that incorporates torsion.As it is known, Einstein constructed also
the "Teleparrell Equivalent of General Relativity"
(TEGR) in which the gravitational field
is described by the torsion tensor and not by the curvature one \cite{KH,JW}. In this approach, the Lagrangian of the modified gravity theory is derived from contractions of the torsion tensor, similar to how the Einstein-Hilbert Lagrangian is derived from contractions of the curvature (Riemann) tensor. Therefore, rather than beginning with General Relativity, one can initiate the development of modified gravity theories by starting with the Teleparallel Equivalent of General Relativity (TEGR). By extending the torsion scalar T to a general function within the Lagrangian, one can construct the f(T) modified gravity theory\cite{RF,GR}.\\
A new proposal in modified gravity is to use the coupling between torsion and the trace of the energy-momentum tensor, which is known as the $f(T,\mathcal{T})$ theory. The $f(T,\mathcal{T})$ gravity theory was introduced in \cite{Harko/2014}.This newly formulated f(T, T) theory represents a novel approach to modified gravity, distinctly separate from existing constructions relying on torsion or curvature.When applied within a cosmological context, this theory yields intriguing observational outcomes. Notably, it allows for a coherent explanation encompassing the early inflationary stage, the subsequent matter-dominated expansion that doesn't involve acceleration, and the eventual shift into a late-time accelerating phase. Furthermore, the effective dark energy sector within this framework can exhibit characteristics ranging from quintessence or phantom-like behavior to potentially crossing the phantom-divide during its evolution. Thus,$f(T,\mathcal{T})$ cosmology offers a unified description of the universe evolution.
Recently the cosmological implications of  $f(T,\mathcal{T})$ was proposed by \cite{AAQID/2023} in which the authors analyzed the cosmological behavior of the deceleration, effective equation of state and total equation of state parameters. However, it is seen that the deceleration parameter depicts the transition from deceleration to acceleration and the effective dark sector shows a quintessence-like evolution. Constraining  $f(T,\mathcal{T})$
 gravity with gravitational baryogenesis was studied in \cite{Swagat}. Also $H_0$ Tension in $f(T,\mathcal{T})$ was studied by \cite{Swagat1} in which the authors gave a new way to alleviate $H_0$ Tension in torsion based modified gravity. The consistency of $f(T,\mathcal{T})$ theory with cosmological data and the required physical conditions for a consistent cosmological theory are yet to be confirmed. However, the coupling of torsion and matter provides additional opportunities for describing the properties of dark energy and identifying the source of the observed acceleration. This theory has been evaluated regarding its reconstruction and stability \cite{Junior/2016,Momeni/2014}, acceleration at late times and during inflationary phases \cite{Harko/2014}, the growth rate of sub-horizon modes\cite{Farrugia/2016}, and also in the context of quark stars\cite{Pace/2017}.

 Two valuable frameworks for characterizing the initial development of the universe are the inflationary scenario \cite{linde/2018,gorb/2011,baff/2021} and bouncing cosmology\cite{db/2015,mv/2008}. Inflationary cosmology is a widely accepted model that describes the very early Universe, shortly after the Big Bang. The earliest inflationary models were developed to address issues associated with primordial singularity and emerged organically from the particles and entropy that existed in the universe. However, these models were incomplete modifications of the big bang theory as they assumed the universe was in a state of thermal equilibrium and homogeneity on large scales prior to inflation. The chaotic inflation proposal resolved this issue. The big bounce, also known as the phoenix universe, is a cyclic cosmological model that oscillates between the big bang and big crunch phases. This theory was widely accepted until the theory of inflation emerged as a resolution to the horizon problem, which was prompted by observations showing the large-scale structure of the universe. The authors of the study \cite{s/2008} explored the future evolution of $f(R)$ gravity models, ensuring consistency with local tests and unifying the history of universe expansion. In \cite{KB}, the authors examined the realization of inflation in the context of unimodular $f(T)$ gravity, which is a modification of teleparallel gravity. The study concluded that inflationary unimodular $f(T)$ gravity is a successful candidate for describing the early universe.
 
 The article aims to examine the generation of a viable inflationary era in  $f (T,\mathcal{T})$ gravity, which is a torsion matter coupling gravity. The focus is on evaluating the viability of each inflationary scenario by assessing the inflationary observational indices such as the tensor-to-scalar ratio (r), the scalar spectral index ($n_s$), the running ($\alpha_s$) of the spectral index, and the tensor spectral ($n_T$). The observational indices will be assessed using conventional techniques, and we will examine whether the resulting cosmologies comply with the most recent Planck and BICEP2/Keck Array data \cite{PA/2002,PA/2016}. To conduct our research, we have examined a well-known model $f(T,\mathcal{T}) =\alpha \mathcal{T}+\beta T^2$  where $\alpha$  and $\beta$  are free parameters. The letter is organized as follows: Section \ref{sec2} provides a concise overview of the field equations in $f (T,\mathcal{T})$ gravity. Section \ref{sec3} is dedicated to investigating the potential for slow-roll inflation using  $f (T,\mathcal{T})$ gravity. Our findings are summarized in the concluding remarks in Section \ref{sec4}.

\section{ Brief review in $f (T,\mathcal{T})$ gravity} \label{sec2}
This section provides the basic principles necessary for rebuilding the theories of gravity, specifically in  $f(T)$ and $f(T,\mathcal{T})$.

To define the torsion-based theory, a new connection called the Weitzenb\"ock connection \cite{Aldrovandi/2013} is required. This connection is expressed as  $\tilde{\Gamma}^{\alpha}_{\mu \nu}= e_{a}^{\,\, \alpha} \partial_{\nu} e^{a}_{\,\,\mu}$. These tetrads are related to the metric tensor $g_{\mu \nu}$ at every point x on the spacetime manifold as
\begin{equation}
\label{1}
g_{\mu \nu}(x)= e^{a}_{\,\, \mu}(x) e^{b}_{\,\, \nu}(x) \eta_{ab}.
\end{equation}

Here, $\eta_{ab}= diag(1,-1,-1,-1)$ is the Minkowski metric tensor.
As a result, the torsion tensor that characterizes the gravitational field can be written as:
\begin{equation}
\label{2}
T^{\alpha}_{\,\,\mu \nu}= \Gamma^{\alpha}_{\,\,\nu \mu}- \Gamma^{\,\,\alpha}_{\mu \nu}= e_{a}^{\,\,\alpha}\left(\partial_{\mu} e^{a}_{\,\,\nu}-\partial_{\nu} e^{a}_{\,\,\mu} \right).
\end{equation}
The components of the torsion tensor are used to define both the contortion and superpotential tensor
\begin{eqnarray}
\label{3}
K^{\mu \nu}_{\,\, \alpha} &=& -\frac{1}{2}\left(T^{\mu \nu}_{\,\, \alpha} - T^{\nu \mu}_{\alpha} - T_{\alpha}^{\,\, \mu \nu}   \right),\\
\label{4}
S_{\alpha}^{\,\, \mu \nu} &=& \frac{1}{2}\left(K^{\mu \nu}_{\,\, \alpha} + \delta^{\mu}_{\alpha} T^{\lambda \mu}_{\,\, \lambda}- \delta^{\nu}_{\alpha} T^{\lambda \mu}_{\,\, \lambda}   \right).
\end{eqnarray}
Using equations \eqref{2} and \eqref{4}, one obtains the torsion scalar \cite{Harko/2014, Maluf/2013, Cai/2016}
\begin{equation}
\label{5}
T = S_{\alpha}^{\,\, \mu \nu} T^{\alpha}_{\,\,\mu \nu}= \frac{1}{2} T^{\alpha \mu\nu} T_{\alpha \mu\nu} + \frac{1}{2} T^{\alpha \mu\nu} T_{ \nu\mu \alpha} - T_{\alpha \mu}^{\,\, \,\, \alpha} T^{\nu \mu}_{\,\,\,\, \nu} .
\end{equation}

The action for teleparallel gravity can be defined as follows:
\begin{equation}
\label{6}
S= \int d^4 x\, e\, [ T+ \mathcal{L}_{m}],
\end{equation}
 where $e$ represents the determinant of the tetrad field $e^{a}_{\mu}$, which is equal to the square root of the metric determinant $-g$, and $\mathcal{L}{m}$ is the Lagrangian density of matter. Additionally, the $T$ in teleparallel gravity can be extended to $T+f(T)$, which is known as $f(T)$ gravity. Furthermore, this can be further generalized to become a function of both the torsion scalar and the trace of the energy-momentum tensor $\mathcal{T}$, resulting in $f(T,\mathcal{T})$ gravity.
 
The action for $f(T,\mathcal{T})$ gravity is defined as follows
\begin{equation}
\label{7}
    S =\frac{1}{16\pi G} \int d^4x\, e[T+f(T,\mathcal{T})]+\int d^4x\, e\, \mathcal{L}_m .
\end{equation}
\textbf{$L_m=-P$ where $P$ is the total pressure.
\\}

The field equations for $f(T,\mathcal{T})$ gravity can be obtained by varying the action with respect to the vierbeins.
\begin{multline}
\label{8}
(1+f_T)\left[e^{-1}\partial_\mu (e e_{a}^{\,\,\alpha} S_{\alpha}^{\,\, \lambda \mu} )-e_{a}^{\,\, \alpha} T^{\mu}_{\nu \alpha} S_{\mu}^{\,\, \nu \lambda} \right] +  e_{a}^{\,\, \lambda}\left(\frac{f+T}{4}\right)+\\
 \left(f_{TT} \partial_{\mu}T+ f_{T\mathcal{T}}\partial_{\mu}\mathcal{T} \right) e_{a}^{\,\, \alpha} S_{\alpha}^{\,\, \lambda \mu} - f_{\mathcal{T}} \left(\frac{e_{a}^{\,\,\alpha} \overset{em}{T}_{\alpha}^{\,\, \lambda} + p_{m} e_{a}^{\,\, \lambda}}{2}\right)= \\
 4 \pi G e_{a}^{\,\,\alpha} \overset{em}{T}_{\alpha}^{\,\, \lambda}.
\end{multline}
where $f_\mathcal{T}={\partial f}/{\partial \mathcal{T}} $, $ f_{T \mathcal{T}}={\partial^2 f}/{\partial T\partial \mathcal{T}}$, and $\overset{em}{T}_{\alpha}^{\,\, \lambda}$ is the energy-momentum tensor.\\

To apply the aforementioned theory in a cosmological framework and obtain modified Friedman equations, we can utilize the flat FRW metric as usual. The FRW metric is given by
\begin{equation}
\label{9}
 ds^{2}=dt^{2}-a(t)^{2} \delta_{ij}dx^{i} dx^{j}, 
\end{equation}
where $a(t)$ is the scale factor. Further, \eqref{8} give rise to modified Friedmann equations:
\begin{equation}
\label{10}
H^2 =\frac{8\pi G}{3}\rho_m - \frac{1}{6}\left(f+12H^2f_T \right)+f_\mathcal{T}\left(\frac{\rho_m+p_m}{3} \right),
\end{equation}
\begin{multline} 
\label{11}
\dot{H}= -4\pi G(\rho_m+p_m)-\dot{H}(f_T-12H^2 f_{T \mathcal{T}})\\-H(\dot{\rho_m}-3\dot{p_m}) f_{T \mathcal{T }} - f_\mathcal{T}\left(\frac{\rho_m+p_m}{2} \right).
\end{multline}
Here, (dot represents first order derivative with respect to $t$) and $\mathcal{T}=\rho_m-3p_m $ in the above equation is true for the perfect matter fluid.

Comparing the modified Friedmann equations \eqref{10} and \eqref{11} to General Relativity equations 
\begin{eqnarray}
\label{12}
H^2 &=& \frac{8 \pi G}{3}\left(\rho_m + \rho_{eff}\right),\\
\label{13}
\dot{H} &=& - 4\pi G \left(\rho_m + p_m + \rho_{eff}+p_{eff}\right).
\end{eqnarray}
We obtain
\begin{equation} 
 \label{14} 
 \rho_{eff} =-\frac{1}{16\pi G}[f+12f_T H^2-2f_\mathcal{T}(\rho_m +p_m)],
 \end{equation}
 \begin{multline} 
 \label{15} 
 p_{eff} = \frac{1}{16\pi G}[f+12f_T H^2-2f_\mathcal{T}(\rho_m +p_m)]+\\
(\rho_m+p_m)+\left[\frac{(1+\frac{f_T}{8\pi G})} {1+f_T 12H^2 f_{TT}+H(\frac{d\rho_m}{dH})(1-3{c_{s}}^2)f_{T \mathcal{T}}} -1\right]
 \end{multline}
Here, $c_s$ is the speed of light. The conservation equation involving the
effective energy and pressure reads as
 \begin{equation}\label{16a}
\dot{\rho}_{\text{eff}} + \dot{\rho}_{m} + 3H(\rho_{m} + \rho_{\text{eff}} + p_{m} + p_{\text{eff}}) = 0
\end{equation}.
Hence, in the current model, the conservation of effective dark energy alone does not hold, and there exists an effective interaction between dark energy and regular matter, allowing for the potential exchange of energy between the two components.
 
\section{Slow-Roll Inflation with  $f(T,\mathcal{T})$ gravity} \label{sec3}
Slow-roll conditions are a set of requirements that inflationary models in cosmology must satisfy to produce a period of exponential expansion in the early universe. Slow-roll conditions have important implications for theories of gravity in cosmology. In particular, they can be used to test and constrain alternative theories of gravity, such as modified gravity models that attempt to explain the observed accelerated expansion of the universe without the need for dark energy. By studying the properties of the cosmic microwave background radiation and large-scale structure, cosmologists can place constraints on the parameters of these theories and rule out those that violate the slow-roll conditions. The slow-roll conditions are typically expressed in terms of the potential energy and the kinetic energy of the inflaton field, as well as its derivatives with respect to time. In particular, for a scalar field $\phi$ with potential energy $V(\phi)$, the slow-roll conditions are.

In every inflationary scenario, one needs to calculate the values of various inflation-related observables,such as the tensor-to-scalar ratio $r$, the scalar spectral index $n_S$, the running of the spectral index $\alpha_s$, and the tensor spectral $n_T$ .Tensor-to-scalar ratio is defined as the ratio of the amplitude of tensor perturbations (primordial gravitational waves) to the amplitude of scalar perturbations.The scalar spectral index $n_s$ describes how the clumpiness of stuff varied on various scales just after cosmic inflation.It is an important parameter describing the nature of primordial density perturbations.  In principle, the
calculation of the above observables requires a detailed
and lengthy perturbation analysis. However, one can bypass this procedure by transforming the given scenario
to the Einstein frame, where all the inflation information
is encoded in the (effective) scalar potential $V(\phi)$, defining the slow-roll parameters $\epsilon$, $\eta$ and $\xi$ in terms of this
potential and its derivatives \cite{JAE,MA}.
\begin{equation}
    \epsilon \equiv \left( \frac{M}{2} \right)^2 \left( \frac{1}{V} \frac{dV}{d\phi} \right)^2
\end{equation}
\begin{equation}
    \eta \equiv \frac{M^2}{p V} \left( \frac{d^2V}{d\phi^2} \right)
\end{equation}
\begin{equation}
    \xi^2 \equiv \frac{M^4}{pV^2} \left( \frac{dV}{d\phi} \right) \left( \frac{d^3V}{d\phi^3} \right)
\end{equation}

The slow-roll parameter $\epsilon$ must be much smaller than unity:
$$\epsilon = \frac{1}{2}\left(\frac{V'(\phi)}{V(\phi)}\right)^2 \ll 1,$$
where $V'(\phi)$ is the derivative of the potential energy with respect to $\phi$.\\
The second slow-roll parameter $\eta$ must also be much smaller than unity:
$$\eta = \frac{V''(\phi)}{V(\phi)} \ll 1,$$
where $V''(\phi)$ is the second derivative of the potential energy with respect to $\phi$.

In this part, we are considering that the conditions for slow-roll inflation are met within the $f(T,\mathcal{T})$ framework, and these conditions are stated in relation to the parameter $H$ as
\begin{eqnarray}
    \label{16}
    \frac{\dot{H}}{H^2} \leq1 , \\ 
    \label{17}
    \frac{\ddot{H}}{H \dot{H}} \leq 1.
\end{eqnarray}

To discuss the possibility of inflation in a theory of gravity that includes a coupling between torsion  and trace, let's consider the following Lagrangian as an example.

We have a function$f(T,\mathcal{T)}=\alpha\mathcal{T}+\beta T^2=\alpha \rho_m +\beta T^2 =\alpha \rho_m+ \gamma H^4 $, where  $\alpha $ and $\gamma=36\beta$ are constants. To simplify, we use $8\pi G = c = 1$. This model is a deviation from General Relativity within the framework of $f(T,\mathcal{T)}$. When $\alpha = 0$ 
 , the model behaves as a power-law cosmology in $f(T)$ theory  \cite{Capozziello/2011}. In this case, we get $f_T=\frac{\gamma T}{18}$, $f_{TT} =\frac{\gamma}{18}$ ,$f_\mathcal{T}=\alpha$,  $f_{T \mathcal{T}}=0$.\\
Using these equations along with equations \eqref{10} \& \eqref{11}, we can derive the following:
\begin{eqnarray}
\label{18}
\rho_{m} &=& \frac{3\left(1-\frac{\gamma H^2}{2} \right)}{1+\frac{\alpha}{2}} H^2, \\ 
\label{19}
\dot{H} &=& -\frac{3(1+\alpha) \left(1-\frac{\gamma H^2}{2}\right)}{(\alpha +2) (1-\gamma H^2)} H^2.
\end{eqnarray}
The deceleration parameter, denoted as $q$, is defined as $q= -\frac{\dot{H}}{H^2}-1$. This progression aligns with the recent behavior of the universe, characterized by three distinct stages: an initial decelerating phase, a subsequent phase of accelerating expansion, and a late-time acceleration phase. For our model, the deceleration parameter $q$ is given as:
\begin{equation}
    \label{20}
 q =\frac{3(1+\alpha)\left(1-\frac{\gamma H^2}{2}\right)}{(\alpha+2)(1-\gamma H^2)}-1.
 \end{equation}


 
Moreover, the effective dark energy density and pressure from equations \eqref{14} and \eqref{13} can be obtained as
\begin{eqnarray}
\label{21}
    \rho_{eff} &=&\frac{3H^2(\alpha+\gamma H^2)}{\alpha+2},\\
    \label{22}
    p_{eff}&=&-\frac{3H^2(\alpha+\gamma H^2)}{(\alpha+2)(\gamma H^2-1)}.
    \end{eqnarray}
which gives
\begin{equation}
    \label{23} 
    \omega_{eff}=\frac{1}{1-\gamma H^2}.
\end{equation}

Applying the slow roll conditions and using the above equation in equation \eqref{16a}, we have
\begin{eqnarray} 
\label{24}
2\dot{H}\gamma+3k_1(\gamma*H^2-2)=0 ,\\  
\label{25}
H(t)=p \text{tan}(q_1t) ,\\
\label{26}
H(t)=p\text{tan}\mathcal{T}_1,
\end{eqnarray}
where 
\begin{eqnarray} \label{27}
p=\frac{\sqrt{2}}{\sqrt{-\gamma}},\,\,\, k_1 = \frac{(\alpha+1)}{(\alpha+2)},\\
q_1=\sqrt{2}\sqrt{-\gamma}, \,\,\, q_1\,t=\mathcal{T}_1
\end{eqnarray}
In the slow roll regime,  \eqref{23}  yields
\begin{equation}
 \label{29}
\omega_eff= \frac{1}{1-\gamma*p^2*\text{tan}^2q_1t}\\
\end{equation}
The inflationary model offers a coherent and effective explanation for the rapid expansion of the universe and the resulting cosmological perturbations that cause its anisotropy. For this, we verified the profile of the deceleration parameter $q(t)$ and found that at a very early evolution process, it presents a rapid expansion and then converges to the de Sitter expansion in the late-time evolution. 
According to the Planck results, the value of $n_s$ is estimated to be $0.968\pm0.006$ (with $68\%$ confidence level), the value of $r$ is less than 0.11 (with $95\%$ confidence level), and the value of $\alpha_s$ is estimated to be $-0.003\pm 0.007$ (with $68\%$ confidence level). These parameters are derived from the slow-roll parameters \cite{FR,KB}.
\begin{eqnarray}\label{30}
 \epsilon_1 \equiv -\frac{\dot{H}}{H^2},\\
 \label{31}
 \epsilon_2 \equiv \frac{\ddot{H}}{H\dot{H}} - \frac{2\dot{H}}{H^2},\\
 \label{32}
 \epsilon_3 \equiv \left(H\ddot{H} - 2\dot{H}^2\right)^{-1},
 \end{eqnarray}
and rewritten as 
\begin{eqnarray}\label{33}
 r\approx 16\epsilon_1,\\ 
 \label{34}
 n_s \approx 1 - 2\epsilon_1 - 2\epsilon_2, \\
 \label{35}
 n_s \approx  - 2\epsilon_1\epsilon_2 - \epsilon_2\epsilon_3, \\
 \label{36}
 n_T \approx -2\epsilon_1.
\end{eqnarray}

Now, we can rewrite the slow-roll parameters as follow;
\begin{eqnarray}\label{37}
 r=-\frac{16 q_1 \csc^2{\mathcal{T}_1}}{p}, \\
 \label{38}
 n_s=1 + \frac{2q_1(-1+\cot^2{\mathcal{T}_1})}{p}, \\
 \label{39}
\alpha_s=-\frac{2q_1^2(1-2pq_1+\cos{2\mathcal{T}_1})\cot^2{\mathcal{T}_1}\csc^4{\mathcal{T}_1}}{p^2(pq_1-\cot^2{\mathcal{T}_1})},\\
\label{40}
 n_T= -\frac{2q_1}{p}\csc^2{\mathcal{T}_1}. 
\end{eqnarray}
We can estimate the number of e-folds\\
\begin{equation}\label{41}
N = \ln \frac{a_f}{a_i} = \int_{t_i}^{t_f} H(t)dt.
\end{equation}
Hence 
where $a_i = a(t=t_i)$ is the initial value of the scale factor $a$ at the beginning of inflation $t_i$, and $a_f = a(t=t_f)$ is its final value at the end of inflation $t_f$.\\
\begin{equation}
\label{42}
N=\frac{p}{q_1} \ln \left[ \frac{\cos(q_1t_f)}{\cos(q_1t_i)} \right].
\end{equation}
Assuming that the Hubble parameter $H(t)$ can be expanded as a series around $\mathcal{T}_1= 0$, we can obtain the second-order approximation.\\
\begin{equation}
   \label{43}
    H(t) \approx \ \ p \mathcal{T}_1 + \mathcal{O}(\mathcal{T}^2_1)
\end{equation}
and the slow-roll parameters become
\begin{eqnarray}
\label{44}
\varepsilon_1 \approx \ \ -\frac{1}{\mathcal{T}^2_1},\\
\label{45}
\varepsilon_2\approx \ \ -2\frac{1}{p\mathcal{T}^2_1},\\
\label{46}
\varepsilon_3\approx \ \ -2\frac{1}{p\mathcal{T}^2_1}.
\end{eqnarray}
Hence in this scenario, we have\\
\begin{eqnarray}
\label{47}
r\approx \ \ -16\frac{1}{p\mathcal{T}_1},\\
\label{48}
n_s=1,\\
\label{49}
\alpha_s \approx \ \ -\frac{8}{p^2 \mathcal{T}^2_1},\\
\label{50}
n_T  \approx \ \ -2 \frac{1}{p\mathcal{T}^2_1}.
\end{eqnarray}
According to Hubble parameter $h(t)$, the e-folding number is  given by 
\begin{equation}\label{51}
    N=\frac{pq_1(t_f^2-t_i^2)}{2}.
\end{equation}
These statements indicate that when the Hubble parameter $H(t)$ can be expanded in a series around $\mathcal{T} = 0$, which occurs for low values of $\mathcal{T}$, the inflationary parameters have large values when the slow-roll conditions are satisfied when the slow-roll conditions are met.\\
In order to evaluate the feasibility of our model, we provide the numerical outcomes for the different inflation parameters given by \eqref{37}, \eqref{38}, \eqref{39} by comparing theory results with observational data coming from PLANK 2015 AND BICEP2/LECK-Array-data \cite{PA/2002,PA/2016}.
\begin{figure}
\centering
\begin{subfigure}[H]{0.5\textwidth}
\includegraphics[scale=0.6]{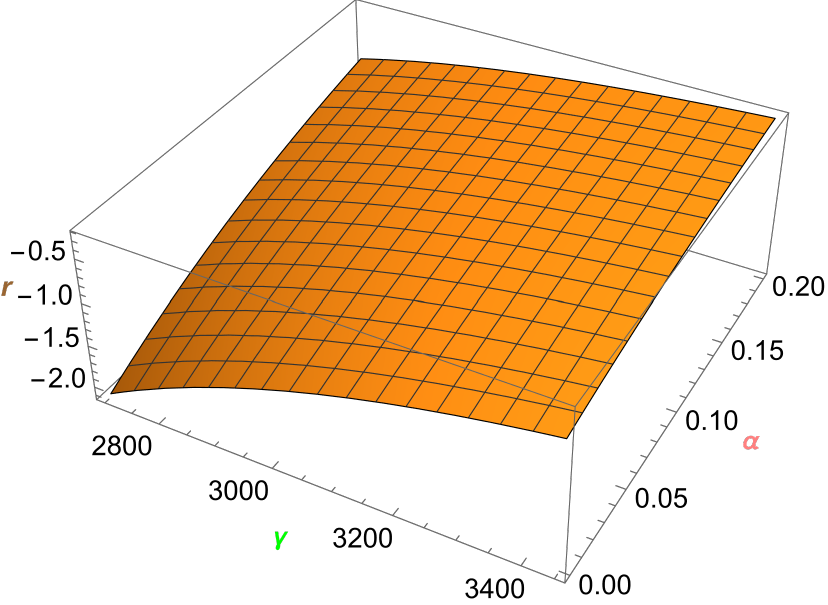}
\end{subfigure}\\
\begin{subfigure}[H]{0.5\textwidth}
\includegraphics[scale=0.7]{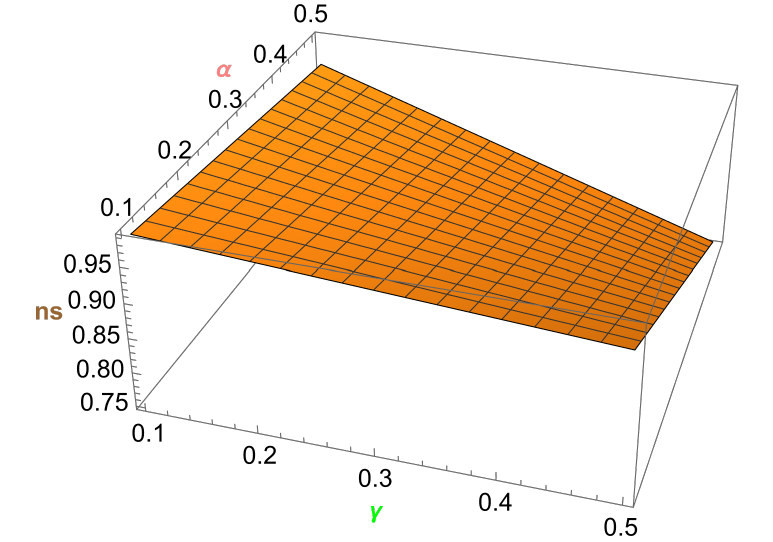}
\end{subfigure}\\
\begin{subfigure}[H]{0.5\textwidth}
\includegraphics[scale=0.68]{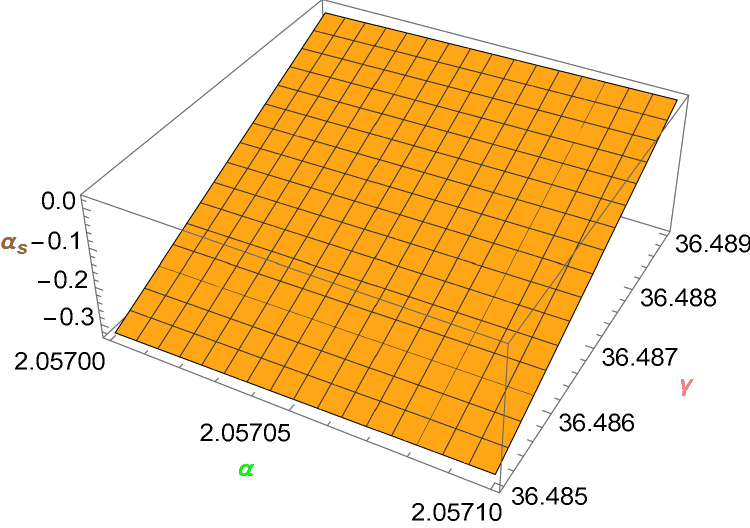}
\end{subfigure}
\caption{Profiles of tensor-to-scalar ratio $(r)$, scalar spectral index $(n_s)$, and running spectral index $(\alpha_s)$ with varying model parameters $\alpha$ and $\gamma$.}
\end{figure}

{The given graphs illustrate how the inflation parameters of model $f(T,\mathcal{T})$ change over time $t = 0.1$s. The first graph shows the tensor-to-scalar ratio $r$ for $\alpha$ values ranging from $0.0$ to $0.2$, while the second graph displays the scalar spectral index $ns$ for $\alpha$ values between $0.1$ to $0.5$. The final graph demonstrates the trend of the running of the  spectral index $\alpha_s$. The evaluation of the curves in the parametric diagram suggests that the tensor-to-scalar ratio $(r)$, scalar spectral index $(ns)$, and running  of the spectral index $(\alpha_s)$ values are consistent with the observed data.}

\section{Conclusion} \label{sec4}
This article investigates the inflationary scenario in torsion-trace coupling gravity, which involves a Lagrangian density derived from a function called $f(T,\mathcal{T})$, which makes use of the torsion $T$ and the trace of the energy-momentum tensor $(\mathcal{T})$.We found evolution of deceleration parameter, explicitly experiencing a change from
a deceleration to acceleration, capable of explaining the
late-time universe. The study focuses on a universe containing ordinary matter and dark energy, and a primary differential equation in the Hubble parameter is derived. Slow-roll conditions are applied for a specific $f(T,\mathcal{T})$ model, and various measurements related to inflationary phenomena are evaluated, including the ratio of tensor to scalar perturbations $(r)$, the spectral index of scalar perturbations $(ns)$, the running of the  spectral index $(\alpha_s)$, the tensor spectral $n_T$, and the number of e-folds parameter. The numerical findings are consistent with observational data. So far, many studies have been done in the background of the torsion-based modified theory focusing on the current acceleration scenario of the universe, and its' late-time acceleration \cite{Harko/2014, AAQID/2023, dsg/2016}. Whereas in this letter, we aimed to study the early-inflationary scenario and successfully presented it. \{We have not only successfully presented the slow-roll inflationary scenario but also presented a way to constrain the $f(T,\mathcal{T})$ cosmological model's parameters. We also constrained the model parameters for the desirable results and discussed them in the last of Section \ref{sec3}. We hope this study will shed some light on a new direction in torsion-based modified gravity cosmology.

{In the near future, one can study more generalized models to put constraints on the parameters and also may extend this study to explore some possible cosmological scenarios. Upcoming surveys are expected to offer a more precise depiction of the universe, leading to a reduction in the number of potential candidates that could provide a more comprehensive explanation for the inflationary period.}

\textbf{Data availability} There are no new data associated with this article.

\acknowledgments  Aaqid Bhat expresses gratitude to the BITS-Pilani, Hyderabad campus, India, for granting him a Junior Research Fellowship. SM acknowledges Transilvania University of Brasov for providing Postdoctoral fellowship. PKS acknowledges Science and Engineering Research Board, Department of Science and Technology, Government of India for financial support to carry out Research project No.: CRG/2022/001847 and IUCAA, Pune, India for providing support through the visiting Associateship program. We are very much grateful to the honorable referees and to the editor for the illuminating suggestions that have significantly improved our work in terms
of research quality, and presentation.
 

\end{document}